\documentclass[11pt,a4paper]{article}
\usepackage{jheppub,dsfont}
\usepackage{epstopdf}

\newcommand{\cc}[2]{c\genfrac{[}{]}{0pt}{}{#1}{#2}}

\newcommand{\nn}{\nonumber}

\notoctrue%
\makeatletter
\renewcommand\maketitle{\pagestyle{empty}
\thispagestyle{titlepage}
\setcounter{page}{0}
\noindent
\@preprint\par
\afterLogoSpace
\if!\@subheader!\else\noindent{\trfont{\@subheader}}\fi
\afterSubheaderSpace
\if!\@proceeding!\else\noindent{\sc\@proceeding}\fi
\afterProceedingsSpace
{\LARGE\flushleft\sffamily\bfseries\@title\par}
\afterTitleSpace
\hrule height 0.2\p@%
\afterRuleSpace
\if!\@collaboration!\else
{\Large\bfseries\sffamily\raggedright\@collaboration}\par
\afterCollaborationSpace
\fi
\if!\@collaborationImg!\else
{\normalsize\bfseries\sffamily\raggedright\@collaborationImg}\par
\afterCollaborationImgSpace
\fi
{\bfseries\raggedright\sffamily\the\auth@toks\par}
\afterAuthorSpace
\ifaffil\begin{list}{}{%
\setlength{\leftmargin}{0.28cm}%
\setlength{\labelsep}{0pt}%
\setlength{\itemsep}{\affiliationsSep}%
\setlength{\topsep}{-\parskip}}
\itshape\small%
\the\affil@toks
\end{list}\fi
\afterAffiliationSpace
\ifemailadd 
\noindent\hspace{0.28cm}\begin{minipage}[l]{.9\textwidth}
\begin{flushleft}
\textit{E-mail:} \the\email@toks
\end{flushleft}
\end{minipage}
\else 
\PackageWarningNoLine{\jname}{E-mails are missing.\MessageBreak Please use \protect\emailAdd\space macro to provide e-mails.}
\fi
\afterEmailSpace
\if!\@xtum!\else\noindent{\@xtum}\afterXtumSpace\fi
\if!\@abstract!\else\noindent{\renewcommand\baselinestretch{.9}\textsc{Abstract:}}\ \@abstract\afterAbstractSpace\fi
\if!\@keywords!\else\noindent{\textsc{Keywords:}} \@keywords\afterKeywordsSpace\fi
\if!\@arxivnumber!\else\noindent{\textsc{ArXiv ePrint:}} \href{http://arxiv.org/abs/\@arxivnumber}{\@arxivnumber}\afterArxivSpace\fi
\if!\@dedicated!\else\vbox{\small\it\raggedleft\@dedicated}\afterDedicatedSpace\fi
\ifnotoc\else
\iftoccontinuous\else\newpage\fi
\beforetochook\hrule
\tableofcontents
\afterTocSpace
\hrule
\afterTocRuleSpace
\fi
\setcounter{footnote}{0}
\pagestyle{myplain}\pagenumbering{arabic}
\vfill
}
\makeatother
\title{Top quark mass coupling and classification of weakly-coupled heterotic superstring vacua}
\author{J. Rizos}
\affiliation{Physics Department, University of Ioannina, GR45110 Greece}

\emailAdd{irizos@uoi.gr}
\abstract{The quest for the Standard Model among the huge number of string vacua is usually based on a set of phenomenological criteria
related to the massless spectrum of string models. In this work we study criteria associated with interactions in the effective low energy theory and in particular with the presence of the coupling that provides mass to the top quark. 
 Working in the context of the Free Fermionic Formulation of the heterotic superstring, we demonstrate that, in a big class of phenomenologically promising $Z_2\times Z_2$ compactifications, these criteria can be expressed entirely in terms of the generalised GSO projection coefficients entering the definition of the models. They are shown to be very efficient in identifying phenomenologically viable vacua, especially in the framework of computer-based search, as they are met by approximately one every $10^4$ models.
We apply our results in the investigation of a class of supersymmetric Pati--Salam vacua, comprising $10^{16}$ configurations, and show that when combined with other phenomenological requirements they lead to a relatively small set of about $10^7$ Standard Model compatible models that can be fully classified.
}

\begin{document} 
\maketitle
\section{Introduction}
One of the main features of string theory is that it accommodates gauge theory that can comprise the Standard Model (SM).
However, string theory in four dimensions, admits an enormous number of vacua and in this context the SM 
appears as just one possibility among others. Recently, considerable effort has been made to classify  
these vacua, from the point of view of low energy phenomenology, in various formulations \cite{tyiim,hetclass,cformalism}.
For this purpose, some phenomenological criteria are employed, usually related to massless spectrum of the model (e.g., number of 
generations, Higgs doublets, SM singlets, exotics). These criteria are often implemented in a computer programme that can 
scan large sets of string vacua. However, criteria associated with interactions in the effective  low energy  theory,
are rarely used since they entail, model dependent, detailed calculations of string amplitudes.
Therefore, it is worthwhile to study the incorporation of interaction related criteria 
in the quest for the SM among the plethora of string vacua.
As a minimum requirement we consider an interaction term associated with the coupling providing mass to, the heaviest fermion, the top quark.
To this end, we demand the presence of a coupling of the form $Q_L\,u_R\,H_u$ at the (tree-level) low energy string effective action, where  $Q_L$ is the left/right-handed quark doublet, $u_R$ the right handed up-quark singlet and $H_u$ the electroweak Higgs doublet. This direction is also motivated by the recent results regarding the discovery of the Higgs scalar particle.
In order to implement this, we are going to work in the framework of the Free Fermionic
Formulation of the heterotic superstring \cite{fff} and in particular in a phenomenologically  interesting class of supersymmetric $Z_2\times Z_2$
vacua, studied in \cite{cformalism}, that exhibit $SO(10)$ embeddable (observable) gauge group symmetry\footnote{For calculations of the top-quark mass in the framework of Free Fermionic models see \cite{topp}.}.


In the framework of Free Fermionic Formulation a string model
is defined in terms of a set of basis vectors, associated with the parallel transformation properties 
of world-sheet fermionic  degrees of freedom along the two non-contractible loops of the world-sheet torus,
and a set of phases associated with generalised GSO projections (GGSO). 
Following a phenomenology oriented approach,  a new method has been proposed in \cite{cformalism} for systematically analysing large classes of $Z_2\times{Z_2}$ heterotic superstring vacua.
It employs a 
fixed set of 12 basis 
vectors $\{v_1,\dots,v_{12}\}$, symmetric in the internal coordinates, namely
\begin{align}
v_1=\mathds{1}&=\{\psi^\mu,\
\chi^{1,\dots,6},y^{1,\dots,6},\omega^{1,\dots,6}|\bar{y}^{1,\dots,6},
\bar{\omega}^{1,\dots,6},\bar{\eta}^{1,2,3},
\bar{\psi}^{1,\dots,5},\bar{\phi}^{1,\dots,8}\}\nn\\
v_2=S&=\{\psi^\mu,\chi^{1,\dots,6}\}\nn\\
v_{2+i}=e_i&=\{y^{i},\omega^{i}|\bar{y}^i,\bar{\omega}^i\}, \ i=1,\dots,6\nn\\
v_{9}=b_1&=\{\chi^{34},\chi^{56},y^{34},y^{56}|\bar{y}^{34},\bar{y}^{56},
\bar{\eta}^1,\bar{\psi}^{1,\dots,5}\}\label{basis}\\
v_{10}=b_2&=\{\chi^{12},\chi^{56},y^{12},y^{56}|\bar{y}^{12},\bar{y}^{56},
\bar{\eta}^2,\bar{\psi}^{1,\dots,5}\}\nn\\
v_{11}=z_1&=\{\bar{\phi}^{1,\dots,4}\}\nn\\
v_{12}=z_2&=\{\bar{\phi}^{5,\dots,8}\}\nn\ .
\end{align}
Here $\psi^\mu$ , $\chi^I$, ${I=1,\dots,6}$ stand for the fermionic superpartners of the 10-dimensional left-moving coordinates, $y^I,\omega^I/\bar{y}^I,\bar{\omega}^I$, ${I=1,\dots,6}$ correspond to the six internal fermionised left/right coordinates, and $\bar{\psi}^A, A=1,\dots,5$, $\bar{\eta}^\alpha,\alpha=1,2,3$, $\bar{\phi}^k,k=1,\dots,8$ are the 16 additional complex fermions of the right-moving sector. The presence of vector $\mathds{1}$ is required for consistency, $S$ is necessary for space-time supersymmetry, $b_1,b_2$ correspond to the $Z_2\times Z_2$ orbifold twists and $e_i,i=1\dots,6$ represent the orbifold shifts. The last two vectors $z_1,z_2$ 
account for possible ``hidden sector" gauge group breaking that gives rise to a richer model structure. The above 12 basis vectors yield
string models with 
$G=SO(10)\times{U(1)}^3\times{SO(8)}^2$ 
gauge symmetry, apart from special points where we have the appearance of gauge group enhancements. 
Matter multiplets in the spinorial representation of the (observable) $SO(10)$ gauge symmetry, appropriate for accommodating the SM fermions,
arise from the three twisted sectors
\begin{align}
b^{1F}_{\left(p_1,q_1,r_1,s_1\right)}&=S+b_1+p_1\,e_3+q_1\,e_4+r_1\,e_5+s_1\,e_6=&\nonumber\\
&=\left\{\psi^\mu,x^{12},w^3_{p_1},w^4_{q_1},w^5_{r_1},w^6_{s_1};\bar{w}^3_{p_1},\bar{w}^4_{q_1},\bar{w}^5_{r_1},\bar{w}^6_{s_1},\bar{\psi}^{123},\bar{\psi}^{45},\bar{\eta}^1\right\}
\label{sI}\\
b^{2F}_{\left(m_2,n_2,r_2,s_2\right)}&=S+b_2+m_2\,e_1+n_2\,e_2+r_2\,e_5+s_2\,e_6=&\nonumber\\
&=\left\{\psi^\mu,x^{34},w^1_{m_2},w^2_{n_2},w^5_{r_2},w^6_{s_2};
\bar{w}^1_{m_2},\bar{w}^2_{n_2},\bar{w}^5_{r_2},\bar{w}^6_{s_2},\bar{\psi}^{123},\bar{\psi}^{45},
\bar{\eta}^2\right\}
\label{sII}\\
b^{3F}_{\left(m_3,n_3,p_3,q_3\right)}&=S+b_1+b_2+x+m_3\,e_1+n_3\,e_2+p_3\,e_3+q_3\,e_4=&\nonumber\\
&=\left\{\psi^\mu,x^{56},w^1_{m_3},w^2_{n_3},w^3_{p_3},w^4_{q_3};
{\bar{w}^1_{m_3},\bar{w}^2_{n_3},}\bar{w}^3_{p_3},\bar{w}^4_{q_3},\bar{\psi}^{123},\bar{\psi}^{45},\bar{\eta}^3\right\}
\label{sIII}
\end{align}
where $p_i,q_i,r_j,s_j,m_k,n_k\in\{0,1\},\,i=1,3,\,j=1,2,\,k=2,3$ and $x=1+S+\sum_{i=1}^6e_i+\sum_{k=1}^2z_k$.
Here we have introduced the convenient notation
$w^i_k=\left(\begin{array}{c}y^i\\\omega_i\end{array}\right)_{k+1}$,
that is, $w^i_0=y^i$ and $w^i_1=\omega^i$ and similarly for $\bar{w}^i_k$. The above sectors can give rise to $3\times 16=48$ spinorials, 16 from
each plane,
corresponding to the orbifold fixed points. Matter in the vectorial representation comes also from the twisted sectors $b^{I\,F}+x,I=1,2,3$
as well as from the untwisted sector. The superparteners of twisted states come from the sectors $S+b^{I\,F}$ and $S+b^{I\,F}+x$.
One or more additional vectors, involving right-moving fields, can be used to break $SO(10)$ to the SM  gauge symmetry  or some subgroup of $SO(10)$ that contains the SM, e.g. $SU(4)\times SU(2)_L\times SU(2)_R$ , $SU(5)\times U(1)$, but the precise form of these vectors is not
important for our analysis.  

As mentioned earlier, the construction of a consistent string model, in this framework, requires also the introduction  of a generalised GSO projection, related to a set of phases, traditionally denoted by $\cc{v_i}{v_j}$. 
In the case of periodic-antiperiodic boundary conditions vectors \eqref{basis}, these phases are real and can take values $\pm1$. Moreover, 
only the ones with $i>j$ turn out to be independent, so altogether we have $2^{12(12-1)/2}=2^{66}\sim10^{20}$ possible configurations.
The strategy followed in \cite{cformalism}, in order to deal with such large number of models, was to derive analytic formulae 
for the main model characteristics in terms of the  GGSO phases $\cc{v_i}{v_j}, i>j$ and to classify the models with respect to a set of 
phenomenological criteria related to their spectrum.

In this article we study the implementation of string vacuum selection criteria related to  the presence of the top-quark mass coupling at the tree-level superpotential. In the first section, we compute the relevant string amplitudes and derive the necessary constraints. In section two,  we apply our results in the classification of Pati--Salam models \cite{pshm}. In the last section we present our conclusions.

\section{Top mass Yukawa coupling}
In the context of a string model generated by basis \eqref{basis}, enhanced by one or more extra vectors that induce $SO(10)$ gauge symmetry
breaking to a subgroup $G$ of $SO(10)$, (e.g. $SU(4)\times SU(2)_L\times SU(2)_R$ , $SU(5)\times U(1)$),  the candidate top mass Yukawa coupling will have the form
\begin{align}
\lambda_t\,\mathbf{S}^{Q_L}\,\mathbf{S}^{u_R}\,\mathbf{V}^{H_u}
\label{tcc}
\end{align}
where $\mathbf{S}$ is the ``spinorial"  and $\mathbf{V}$ the ``vectorial" representation of $G$. We use the terms ``spinorial/vectorial" to
 denote the representations of $G\subset SO(10)$ that are accommodated in the spinorial/vectorial of $SO(10)$ respectively.
In the sequel, we will continue using this terminology omitting the quotes.
The superscripts indicate the SM fields entering the top mass coupling. 
Since we consider $N=1$  supersymmetric models, the coupling \eqref{tcc} arises from a superpotential interaction of the form
\begin{align}
\lambda_t\,\int d^2\theta\,{\Phi_\mathbf{S}}\,{\Phi_\mathbf{S}}\,{\Phi_\mathbf{V}}
\end{align}
where ${\Phi_\mathbf{S,V}}$ are the associated superfields. In the context of the Free-Fermionic-Formulation of the heterotic superstring the coupling constant $\lambda_t$ is fully calculable and is proportional to the correlation function of the associated vertex operators of the massless string modes 
\begin{align}
\lambda_t\sim \left<S^F_{-1/2}\,S^F_{-1/2},V^B_{-1}\right>
\label{stt}
\end{align}
where the superscript $F,B$ denotes the fermionic or bosonic part respectively and the subscript indicates the BRST ghost charge \cite{Kalara:1990fb}. 
This correlator factorises  into a product of terms involving: (i) the BRST ghosts;  (ii) the spacetime spin;  (iii) spacetime momentum;  (iv) the bosonised left-moving complex fermions; 
(v) the real left-right Ising fermions  and  (vi) the local gauge group generating fermions. 
Of particular interest is part (iv) associated with the fermions of the spacetime supersymmetry generating vector $S$, that is, 
$\left\{\chi^{12},\chi^{34},\chi^{56}\right\}\,$, as it leads to some model independent selection rules  for the superpotential \cite{Rizos:1991bm}. It can be demonstrated that the only non-vanishing three point correlation functions, of the type \eqref{stt}, are
\begin{align}
\left<(R)^1(R)^2(R)^3\right> \ ,\ \left<(R)^i(R)^i(NS)\right> \, \ i=1,2,3\ ,\ 
\left<(NS) (NS)(NS)\right> 
\label{att}
\end{align}
where $(R)$ stands for a generic twisted (Ramond) and $(NS)$ for an untwisted (Neveau--Schwartz) field and the superscript indicates
the orbifold plane. In \eqref{att} we have dropped the ghost charge subscripts as well as the fermion/boson superscripts of \eqref{stt}.

An important property of the string models under consideration is that the quarks and leptons, which reside in $SO(10)$ spinorials, arise exclusively from the  twisted sector, while SM Higgs fields, residing in the vectorial representation of $SO(10)$,  can arise both from the twisted and the untwisted sector.
Hence, we have to consider only the first two types of couplings in \eqref{att}. 
Starting from the coupling of the form $\left<(R)^1(R)^2(R)^3\right>$,
we can assume, without loss of generality, that the first spinorial representation, $S^{1F}_{\left(p_1,q_1,r_1,s_1\right)}$, of the candidate top Yukawa coupling in \eqref{stt} arises from the first obifold plane and particularly from the sector $b^{1F}_{\left(p_1,q_1,r_1,s_1\right)}$, the second spinorial $S^{2F}_{\left(m_2,n_2,r_2,s_2\right)}$ from the second plane sector $b^{2F}_{\left(m_2,n_2,r_2,s_2\right)}$ and the vectorial representation $V^{1B}_{\left(m_3,n_3,p_3,q_3\right)}$ coming from the 
third plane sector $b^{3B}_{\left(m_3,n_3,p_3,q_3\right)}$, where the first two are given in \eqref{sI}-\eqref{sII}, and 
\begin{align}
b^{3B}_{\left(m_3,n_3,p_3,q_3\right)}&=b_1+b_2+m_3\,e_1+n_3\,e_2+p_3\,e_3+q_3\,e_4=&\nonumber\\
&=\left\{x^{12},x^{34},w^1_{m_3},w^2_{n_3},w^3_{p_3},w^4_{q_3};
{\bar{w}^1_{m_3},\bar{w}^2_{n_3},}\bar{w}^3_{p_3},\bar{w}^4_{q_3},{\bar{\eta}^1,\bar{\eta}^2}\right\}
\label{sIIIb}
\end{align}

Returning to the correlation function \eqref{stt}, we examine the Ising part. 
For a pair of real fermions $\left(f,\bar{f}\right)$,  Ising correlators can involve the following  conformal fields: The left and right fermions $f(z)$, $\bar{f}(z)$, the combination $f(z)\bar{f}(z)$ (energy operator), the spin  $\sigma_+\left(z,\bar{z}\right)$ (order operator) and its dual $\sigma_-\left(z,\bar{z}\right)$ (disorder operator) and of course the
identity operator $I$. Since the vertex operators of the fields, entering the coupling in question, do not involve any real fermion oscillators we consider correlators  involving $\sigma_\pm$ fields. According to \cite{DiFrancesco:1987ez} we have
\begin{align}
\left<\sigma_+\right>=\left<\sigma_-\right>=0\ ,\ \left<\sigma_+\sigma_+\right>=\left<\sigma_-\sigma_-\right>=1\ ,\ \left<\sigma_+\sigma_-\right>=0
\end{align}
These formulae  can be generalized to correlators involving two order/disorder operators associated with some of the 12 pairs of real world-sheet fermions, as follows:
\begin{align}
\left<\sigma_\epsilon\left(w^i_k\right)\,\sigma_\zeta\left(w^j_m\right)\right>=\delta_{ij}\delta_{km}\delta_{\epsilon\zeta}\ ,
\end{align} 
where  $\sigma_\epsilon\left(z^i_k\right)$ denotes the Ising spin field of the $i$-th pair of real fermions  $i=1,\dots6$, of type $k=0,1$ 
($k=0$ for $y^i\bar{y}^i$ and $k=1$ for $\omega^i\bar{\omega}^i$) and subscript $\epsilon=+,-$ ($+$ for order and $-$ for disorder operator).

Applying the above results and using \eqref{sI},\eqref{sII} and \eqref{sIIIb},
we obtain 
\begin{align}
\lambda_t&\sim\left<S^{1F}_{\left(p_1,q_1,r_1,s_1\right)}\,S^{2F}_{\left(m_2,n_2,r_2,s_2\right)}\,V^{3B}_{\left(m_3,n_3,p_3,q_3\right)}\right>\sim\nonumber\\
&\left<\sigma_{\epsilon^1_2}\left(w^1_{m_2}\right)\,\sigma_{\epsilon^1_3}\left(w^1_{m_3}\right)\right>
\left<\sigma_{\epsilon^2_2}\left(w^2_{n_2}\right)\,\sigma_{\epsilon^2_3}\left(w^2_{n_3}\right)\right>
\left<\sigma_{\epsilon^3_1}\left(w^3_{p_1}\right)\,\sigma_{\epsilon^3_3}\left(w^3_{p_3}\right)\right>
\left<\sigma_{\epsilon^4_1}\left(w^4_{q_1}\right)\,\sigma_{\epsilon^4_3}\left(w^4_{q_3}\right)\right>\nn\\
&\hspace{7cm}\times
\left<\sigma_{\epsilon^5_1}\left(w^5_{r_1}\right)\,\sigma_{\epsilon^5_2}\left(w^5_{r_2}\right)\right>
\left<\sigma_{\epsilon^6_1}\left(w^6_{s_1}\right)\,\sigma_{\epsilon^6_2}\left(w^6_{s_2}\right)\right>\nonumber\\
&=\delta_{m_2,m_3}\delta_{n_2,n_3}\delta_{p_1,p_3}\delta_{q_1,q_3}\delta_{r_1,r_2}\delta_{s_1,s_2}
\times
\prod_{i=1,2}\delta_{{\epsilon^i_2},{\epsilon^i_3}}
\prod_{i=3,4}\delta_{{\epsilon^i_1},{\epsilon^i_3}}
\prod_{i=5,6}\delta_{{\epsilon^i_1},{\epsilon^i_2}}
\label{con}
\end{align}
where $\epsilon^i_j$ refers to fermion pair $i$ of plane $j$.
The first conclusion that can be drawn from the above result is 
 that the only non-vanishing couplings are of the form
\begin{align}
\left<S^{1F}_{(p,q,r,s)}\,S^{2F}_{(m,n,r,s)}\,V^{3B}_{(m,n,p,q)}\right>\ \ ,p,q,r,s,m,n={0,1}\ ,
\end{align}
that is, the participating sectors add up to zero
\begin{align}
{b^{1F}}_{(p,q,r,s)}+{b_2^F}_{(m,n,r,s)}+{b_3^B}_{(m,n,p,q)}=0\ .
\end{align}
Therefore, out of the $2^{12}$ candidate sector couplings only $2^6$ are non-vanishing, those
with spinorials coming from the sectors $S+b_1+p\,e_3+q\,e_4+r\,e_5+s\,e_6$, $S+b_2+m\,e_1+n\,e_2+r\,e_5+s\,e_6$ and the vectorial from the sector 
$b_1+b_2+m\,e_1+n\,e_2+p\,e_3+q\,e_4$. Since we need only one such term, to play the role of the top mass coupling, we can redefine our basis vectors
\eqref{basis} to be this specific combination. That is, we set $b_1\to b_1'=b_1+p\,e_3+q\,e_4+r\,e_5+s\,e_6$, $b_2\to b_2'=b_2+m\,e_1+n\,e_2+r\,e_5+s\,e_6$ that consistently yields  $b_1'+b_2'$ 
as the third sector. As a result, we can assume, without loss of generality, that the spinorials of the  top Yukawa coupling arise from sectors $S+b_1,S+b_2$ and the vectorial from $b_1+b_2$.

Another conclusion that can be drawn from \eqref{con} is that the  signs $e^i_j$ have to match. In other words
\begin{align}
{{\epsilon^i_2}={\epsilon^i_3}},i=1,2\ ,\ {\epsilon^i_1}={\epsilon^i_3},{i=3,4}\ ,\ {\epsilon^i_1}={\epsilon^i_2},{i=5,6}\,.
\label{ccc}
\end{align}
 However, $e^i_j$ are associated with the GGSO projections of the shift vectors $e_i,\,i=1,\dots,6$ onto the sectors  $S+b_1,S+b_2$, $b_1+b_2$ as follows:
\begin{align}
\epsilon_1^i=\delta_{S+b_1}\,\cc{S+b_1}{e_i}=\cc{b_1}{e_i}\ ,\ i=3,4,5,6\label{ei}\\
\epsilon_2^j=\delta_{S+b_2}\,\cc{S+b_2}{e_j}=\cc{b_2}{e_j}\ ,\ j=1,2,5,6\\
\epsilon_3^\ell=\delta_{b_1+b_2}\,\cc{b_1+b_2}{e_\ell}=\cc{b_1}{e_\ell}\,\cc{b_2}{e_\ell}\ ,\ \ell=1,2,3,4\label{ef}
\end{align}
Moreover, in order to preserve the states, entering the  coupling in question, from the GGSO projections we must impose
\begin{align}
\delta_{S+b_1}\,\cc{S+b_1}{e_i}=\cc{b_1}{e_i}=+1\ ,\ i=1,2\label{gpi}\\
\delta_{S+b_2}\,\cc{S+b_2}{e_j}=\cc{b_2}{e_j}=+1\ ,\ j=3,4\\
\delta_{b_1+b_2}\,\cc{b_1+b_2}{e_\ell}=\cc{b_1}{e_\ell}\,\cc{b_2}{e_\ell}=+1\ ,\ \ell=5,6\\
\delta_{S+b_k}\,\cc{S+b_k}{z_a}=\cc{b_k}{z_a}=+1\ ,\ a=1,2\,,\,k=1,2\\
\delta_{b_1+b_2}\,\cc{b_1+b_2}{z_a}=\cc{b_1}{z_a}\cc{b_2}{z_a}=+1\ ,\ a=1,2\label{gpf}
\end{align}
This is required since there is no overlap between $e_i,z_k$ and $b_a$, that is, $e_i\cap{S+b_1}= \varnothing,i=1,2,$ $e_i\cap{S+b_2}= \varnothing,i=3,4,$ $e_i\cap{S+b_1+b_2}= \varnothing,i=5,6$
 and also $z_a\cap{S+b_1}=z_a\cap{S+b_2}=z_a\cap{S+b_1+b_2}= \varnothing,a=1,2$.
 
Solving Eqs. \eqref{ccc} , \eqref{gpi}-\eqref{gpf} and using \eqref{ei}-\eqref{ef} we obtain a set of necessary 
conditions for the existence of the top mass coupling \eqref{stt}
\begin{gather}
\cc{b_1}{e_1}=\cc{b_1}{e_2}=\cc{b_2}{e_3}=\cc{b_2}{e_4}=\cc{b_1}{z_1}=\cc{b_1}{z_2}=\cc{b_2}{z_1}=\cc{b_2}{z_2}=+1\nn\\
\cc{b_1}{e_5}\cc{b_2}{e_5}=
\cc{b_1}{e_6}\cc{b_2}{e_6}=+1
\label{fc}
\end{gather}

Let us consider the group part of the correlator \eqref{att}. By construction it is invariant under $SO(10)$, however, we need to examine
the invariance under the $U(1)^3$ part of the gauge symmetry $G$, generated by the right fermionic fields $\bar{\eta}^1,\bar{\eta}^2,\bar{\eta}^3$.
As can be seen from \eqref{sI},\eqref{sII} and \eqref{sIIIb}, $S^{1F}$ carries $\bar{\eta}^1$ charge, $Q^{1F}=\left(q^1_1,0,0\right)$, $S^{2F}$ carries $\bar{\eta}^2$ charge,$Q^{2F}=\left(0,q^2_2,0\right)$, while $V^{3B}$ carries both $\bar{\eta}^1$ and $\bar{\eta}^2$ charges,$Q^{3B}=\left(q^3_1,q^3_2,0\right)$. These charges can be expressed entirely in terms of the GGSO phases as follows
\begin{align}
q^{1}_1&=-\frac{1}{2}c_\psi\cc{S}{S}\cc{b_2}{x}\cc{b_1}{b_2}\\
q^{2}_2&=-\frac{1}{2}c_\psi\cc{S}{S}\cc{b_2}{x}\cc{b_1}{b_2}\\
q^{3}_1&=\frac{1}{2}c_\psi\cc{S}{S}\cc{b_1}{S}\cc{b_1}{b_2}\cc{b_1}{b_1}\\
q^{3}_2&=\frac{1}{2}c_\psi\cc{S}{S}\cc{b_1}{S}\cc{b_1}{b_2}\cc{b_2}{b_2}\,,
\end{align}
where the subscript refers to the corresponding $\bar{\eta}$ field and $c_\psi=\pm1$ is the spacetime chirality of the fermionic component of the
associated fields. In the derivation of this result we have used the GGSO projections of the vectors $S+b_2$, $x$ onto $S+b_1$, 
the projections of $S+b_1$, $x$ onto $S+b_2$, and the projections of $S+b_1$, $S+b_2$ onto $S+b_1+b_2$. Gauge invariance requires
\begin{align}
q^{1}_1+q^3_1=0\Rightarrow \cc{b_2}{S}\cc{b_1}{b_1}=\cc{b_1}{x}\\
q^{2}_2+q^3_2=0\Rightarrow \cc{b_2}{S}\cc{b_2}{b_2}=\cc{b_2}{x}
\end{align}
Taking into account the properties of the GGSO coefficients and \eqref{fc} we find the additional constraints
\begin{align}
\prod_{i=3,4,5,6}\cc{b_1}{e_i}=\prod_{i=1,2,5,6}\cc{b_2}{e_i}=1
\label{fd}
\end{align}

Let us now turn to the analysis of the couplings of the form $\left<(R)^I(R)^I(NS)\right> \,, \ I=1,2,3$ involving a Higgs field from the untwisted sector. The general form of the $SO(10)$ vectorial scalar fields arising from the NS sector is $\left(\chi^{2i-1}+ i\,\chi^{2i}\right)_{1/2}\left|0\right>_{L}\otimes\bar{\eta}^i_{1/2}\bar{\psi}^a_{1/2}\left|0\right>_{R}+\text{c.c.}, i=1,2,3$. Without loss of generality we can take
$I=1$, that is, the two spinorials arise from $S^{1F}_{\left(p_1,q_1,r_1,s_1\right)}$ and $S^{1F'}_{\left(p_1',q_1',r_1',s_1'\right)}$,  in which case the correlators related with $\chi^{1,\dots,6}$ require $i=1$ \cite{Rizos:1991bm}. Since there are no Ising fields involved in the Higgs scalar vertex operator, the  related correlator takes the form
\begin{align}
\lambda_t&\sim\left<S^{1F}_{\left(p_1,q_1,r_1,s_1\right)}S^{1F'}_{\left(p_1',q_1',r_1',s_1'\right)}V^B_{NS}\right>
\nonumber\\
&\sim
\left<\sigma_{\epsilon^3_1}\left(w^3_{p_1}\right)\,\sigma_{\epsilon^{3'}_1}\left(w^3_{p_1}\right)\right>
\left<\sigma_{\epsilon^4_1}\left(w^4_{q_1}\right)\,\sigma_{\epsilon^{4'}_1}\left(w^4_{q_1'}\right)\right>
\left<\sigma_{\epsilon^5_1}\left(w^5_{r_1}\right)\,\sigma_{\epsilon^{5'}_1}\left(w^5_{r_1'}\right)\right>
\left<\sigma_{\epsilon^6_1}\left(w^6_{s_1}\right)\,\sigma_{\epsilon^{6'}_1}\left(w^6_{s_1'}\right)\right>\nonumber\\
&=\delta_{p_1,p_1'}\delta_{q_1,q_1'}\delta_{r_1,r_1'}\delta_{s_1,s_1'}
\times
\prod_{i=3}^6\delta_{{\epsilon^i_1},{\epsilon^{i'}_1}}
\end{align}
Given the fact that only one spinorial is possible for every combination of $p,q,r,s$ (labelling each fixed point) we are led to the 
conclusion that the two spinorial states are actually the same field 
$S^{1F}_{\left(p_1,q_1,r_1,s_1\right)}=S^{1F'}_{\left(p_1',q_1',r_1',s_1'\right)}$. However, no such fermion mass term is allowed in the context
of the SM. Moreover, after examining the subgroups of $SO(10)$ that contain the SM, we find that a mass term of this form is only allowed  by $SU(5)\times{U(1)}$ gauge symmetry, i.e
$\left({\mathbf{10}},\frac{1}{2}\right)^2\left(\mathbf{5},-1\right)$, and can provide masses for the down quarks and leptons, but not
for the up quarks. Hence, couplings of the form $\left<(R)^I(R)^I(NS)\right>$ are not relevant to our analysis.

To summarise, the conditions \eqref{fc},\eqref{fd}, 
 guarantee the presence of the top-quark mass coupling
at the tree-level superpotential. Altogether, they fix 12 GGSO coefficients and reduce the number of acceptable models in this class by a factor of $2^{12}=4096$. Furthermore, some constraints have also to be imposed on the additional GGSO coefficients related to the extra basis vector(s) responsible for the breaking of the $SO(10)$ gauge group factor. For example, in the case of the Pati--Salam models, analysed in the next section, these constraints fix two additional GGSO phases. Thus, in a realistic model, the number of necessary constraints is increased by a factor of four.

\section{Classification of Pati-Salam heterotic superstring vacua}
In this section, we examine the consequences of the top mass coupling selection rule in the classification 
of Pati-Salam (PS) heterotic superstring vacua 
studied in \cite{pshm}. Our purpose is to provide a concrete application of results derived in the previous section and also to examine the effectiveness of the constraints, derived in the previous section, when combined with other phenomenological requirements.

The Pati-Salam model \cite{ps} and its supersymmetric string realisation \cite{pss}, based on the gauge symmetry group $G=SU(4)\times{SU(2)}_L\times{SU(2)}_R$, are of particular phenomenological interest.
The SM particles, residing in the spinorial (${\mathbf{16}}$) of $SO(10)$, are accommodated in PS group representations as follows:
\begin{align}
{\mathbf{16}}=\left\{
\begin{array}{ll}
{F}_L\left({\mathbf{4}},{\mathbf{2}},{\mathbf{1}}\right)&=Q\left({\mathbf{3}},{\mathbf{2}},\frac{1}{6}\right) + L\left({\mathbf{1}},{\mathbf{2}},-\frac{1}{2}\right)\\
\bar{F}_R\left(\bar{\mathbf{4}},{\mathbf{1}},{\mathbf{2}}\right)&=u^c\left(\overline{\mathbf{3}},{\mathbf{1}},-\frac{2}{3}\right)+
d^c\left(\bar{\mathbf{3}},{\mathbf{1}},\frac{1}{3}\right)+e^c\left({\mathbf{1}},{\mathbf{1}},1\right)+
\nu^c\left({\mathbf{1}},{\mathbf{1}},0\right)
\end{array}
\right.\,,
\end{align}
while Higgs doublets together with extra quark triplets, residing in the vectorial ($\mathbf{10}$) of  $SO(10)$, are assigned to
PS group representations as
\begin{align}
{\mathbf{10}}=\left\{
\begin{array}{ll}
D\left({\mathbf{6}},{\mathbf{1}},{\mathbf{1}}\right)&=d\left({\mathbf{3}},{\mathbf{1}},-\frac{1}{3}\right) + \bar{d}\left(\overline{\mathbf{3}},{\mathbf{1}},\frac{1}{3}\right)\\
h\left({\mathbf{1}},{\mathbf{2}},{\mathbf{2}}\right)&=H^d\left({\mathbf{1}},{\mathbf{2}},-\frac{1}{2}\right)+
H^u\left({\mathbf{1}},{\mathbf{2}},\frac{1}{2}\right)
\end{array}
\right.\,.
\end{align}
The $SU(4)\times{SU(2)}_L\times{SU(2)}_R$ gauge group breaks to SM one via VEVs of (at least) one pair
of Higgs fields
$
{H}\left({\mathbf{4}},{\mathbf{1}},{\mathbf{2}}\right)+\bar{H}\left({\overline{\mathbf{4}}},{\mathbf{1}},{\mathbf{2}}\right)   
$. These are accommodated in  pairs of $SO(10)$ spinorials ($\mathbf{16}+\overline{\mathbf{16}}$).
Quark and lepton masses arise, as in the case of $SO(10)$ model, from a single
 superpotential term
\begin{align}
{F}_L\left({\mathbf{4}},{\mathbf{2}},{\mathbf{1}}\right)\,\bar{F}_R\left(\bar{\mathbf{4}},{\mathbf{1}},{\mathbf{2}}\right)\,h\left({\mathbf{1}},{\mathbf{2}},{\mathbf{2}}\right)=Q\,u^c\,H^u+Q\,d^c\,H^d+L\,e^c\,H^d+L\,\nu^c\,H^u 
\,.
\label{tocps}
\end{align}
However, neutrinos stay light through the mixing with additional heavy singlets.

The class of PS string vacua under consideration is generated by the basis \eqref{basis} after the addition of the vector 
\begin{align}
v_{13}=\alpha=\{\bar{\psi}^{45},\bar{\phi}^{1,2}\}\,.
\end{align}
In principle, this class comprises  $2^{78}$ models, however, some of the GGSO phases are fixed from the requirements of $N=1$ supersymmetry,
conventions and symmetries \cite{pshm}. Finally, we are left with $2^{55} \sim 3.6\times 10^{16}$ 
models\footnote{
Extra symmetries of the spectrum have been employed in \cite{pshm} to fix four additional GGSO phases. However, when interactions are taken into
account these phases lead to different models.}. 
The vector $\alpha$ breaks  the gauge symmetry to $SU(4)\times{SU(2)}_L\times{SU(2)}_R\times{U(1)}^3\times{SU(2)}^4\times{SO(8)}$, truncates appropriately the matter spectrum and at the same time gives rise to new particles from the twisted sectors. 

The presence of the top mass coupling \eqref{tocps} in the tree-level effective action requires, 
besides the 12 constraints derived in the previous session,
the survival of the $SU(4)\times{SU(2)}_L\times{SU(2)}_R$ fields involved. That is, the GSO
projections induced by the extra basis vector $\alpha$ must preserve $\left({\mathbf{4}},{\mathbf{2}},{\mathbf{1}}\right)$ 
coming from sector $S+b_1$, $\left(\bar{\mathbf{4}},{\mathbf{1}},{\mathbf{2}}\right)$
coming from $S+b_2$, and $\left({\mathbf{1}},{\mathbf{2}},{\mathbf{2}}\right)$ coming from $b_1+b_2$.
When translated in terms of GGSO phases, these requirements entail
\begin{align}
\cc{b_1}{a}=-\cc{b_2}{a}=+1\,.
\label{ac}
\end{align}
This raises the number of constraints among GGSO phases to 14 and hence the number of phenomenologically acceptable 
vacua in this class is reduced to $2^{41} \sim 2.2\times 10^{12}$. Thus, the top mass coupling constraint turns out to be very 
efficient as it selects one every $2^{14}$ models.

We will now examine the implementation of the above selection rules when combined with other criteria in a computer scan of Pati--Salam vacua. 
In Ref. \cite{pshm}, analytic formulae have been derived expressing some basic phenomenological features of a models in terms of the GGSO phases.
These include: the number of fermion generations ($n_g$), the number of PS breaking Higgs pairs($k_R$), the number of additional vector-like left pairs ($k_L$), the number of bi-doublets which accommodate the SM Higgs doublets ($n_h$), the number of  sextets containing the extra SM triplets ($n_6$), and the number of fractional charge exotic multiplets ($n_e$). 
In this class of models, exotic matter states, that transform as $\left({\mathbf{1}},{\mathbf{2}},{\mathbf{1}}\right)$, $\left({\mathbf{1}},{\mathbf{1}},{\mathbf{2}}\right)$, $\left({\mathbf{4}},{\mathbf{1}},{\mathbf{1}}\right)$,$\left({\bar{\mathbf{4}}},{\mathbf{1}},{\mathbf{1}}\right)$, can arise from the sectors $b^I+a(+z_1)(+S),b^I+a+x(+z_1)(+S),I=1,2,3\,$. Nonetheless, a subclass of models has been discovered, referred as exophobic \cite{pshm},
 whose massless spectrum is free of exotic fractionally charged states. These models are interesting, since light fractional charge exotics are hard to accommodate in the standard cosmological scenario\cite{Langacker:2011db}. 
 
\begin{figure}[!ht]
\centering
\includegraphics[width=\textwidth]{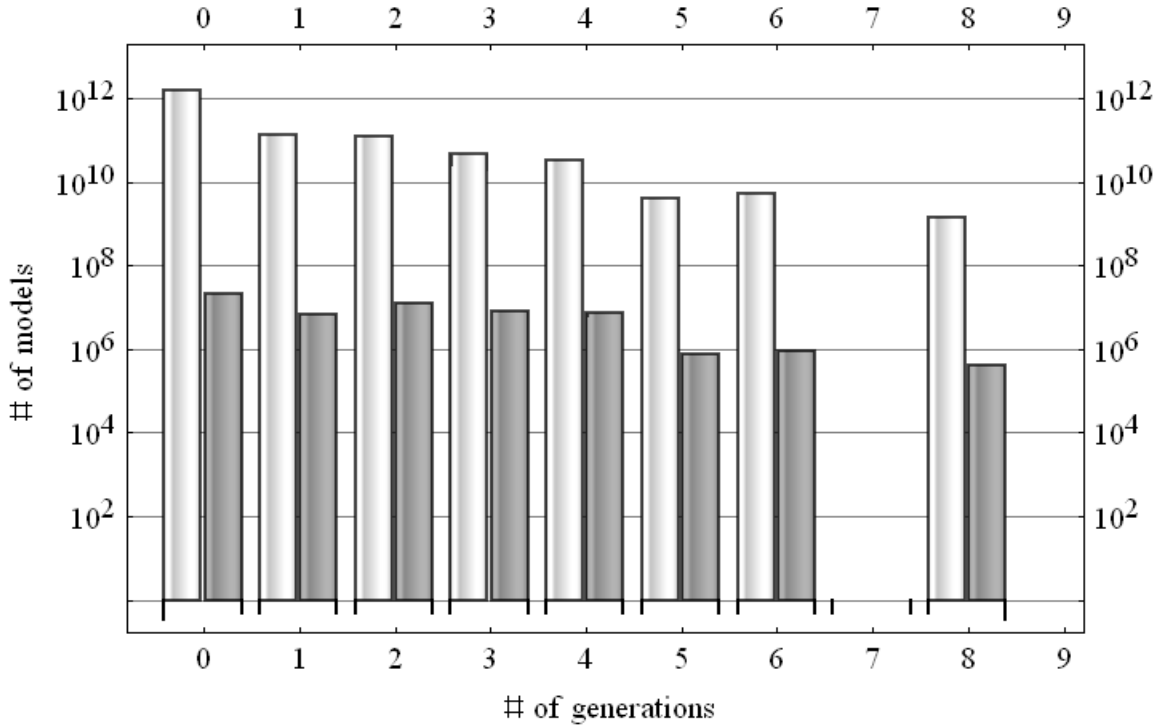}
\caption{\label{gen_distribution}
\it Number of Pati--Salam models, that meet certain phenomenological criteria (see text), versus number of generations, before (light-grey) and after (dark-grey) the application of the top mass coupling constraints}
\end{figure}

Using the analytic formulae of \cite{pshm}, we have calculated the number of PS models 
whose massless spectrum satisfy the following phenomenological requirements: (i) complete fermion generations 
(ii) existence of PS breaking Higgs multiplets ($k_R\ge1$) (iii) existence of SM breaking Higgs doublets ($n_h\ge1$) and (iv)absence of fractional charge exotics($n_e=0$).
The number of acceptable PS models versus the number of generations is depicted in 
Fig. \ref{gen_distribution}. The light-gray columns correspond to the number of models before the application of the top mass coupling constraints and the dark-grey columns correspond to the number of models after the application of the constraints. The former is estimated using a scan over
a random sample of $10^{11}$ models, while the latter is the exact result of a scan over the full space of models.
It is evident that the coupling constraints derived here, reduce the number of acceptable vacua by approximately  four orders of magnitude. Moreover,
as they fix certain GGSO phases, they efficiently reduce the scanning region and allow an
exhaustive scan of phenomenologically interesting vacua. 
It turns out that approximately $10^7$ vacua meet all phenomenological criteria considered above.
Moreover, we have $10^6$ ``minimal models", in the sense that they do not contain any additional vector-like fermion generations ($k_L=0,k_R=1$).
The minimal models fall into four categories, i.e. $(n_h,n_6)\in\left\{(1,3),(3,1),(3,5),(5,3)\right\}$.
The vacua of each category appear to have identical spectra. This raises the question:  
Are all these vacua really different? To answer this question we must first analyse the hidden sector and classify the models using their full spectrum. The analysis here was restricted to the observable sector spectrum. Then we have to compare the models at the level of superpotential couplings. This goes beyond the scope of this article, however, the above results show that this analysis is feasible, at least as far as the spectrum is concerned, due to the relatively small number of acceptable vacua. In fact, some
preliminary calculations show that, even when the hidden sector spectrum is taken into account, we still have a significant model degeneracy.
Another question is whether any of these models will survive a detailed phenomenological analysis. To answer this, we note 
that the model studied in \cite{topm}, although derived in a different context, actually belongs to the above class of ``minimal models" 
($(n_h,n_6)=(1,3)$). This can be easily verified using the replacements $b_1\to b_1'=b_1+b_2+x+e_2+e_4$, $b_2\to b_2'=b_1+e_4+e_5$.
As shown in \cite{topm}, this specific model passes successfully some detailed phenomenological tests including the existence of  
F and D flatness condition solutions, that render the additional triplets super-heavy while keeping light the SM Higgs doublets. 

Summarising, here we have applied the criteria, developed in the previous section,
in the classification of Pati--Salam heterotic supersymmetric vacua and demonstrated that they can be easily implemented and combined
with other phenomenological constraints. We have also identified a relatively small class of models that meet all phenomenological requirements 
and deserve further investigation.

\section{Conclusions}
In this article we have studied the implementation of string vacuum selection criteria, related to the couplings of the effective low energy theory and particularly to the presence of the top-quark mass coupling, $Q_L\,u_R\,H_u$, at tree-level of the superpotential. We have demonstrated that in a big class of string vacua with gauge group $G \subset SO(10)$ the associated correlation function involves only twisted fields and can be explicitly computed in terms of the GGSO phases. Therefore, the requirement of existence  of the top mass coupling is translated into a set of constraints on  the GGSO phases defining the string model. We find that only $1:10^4$ models in this class satisfy these constraints. Thus 
these criteria turn out to be  very efficient in selecting amongst string vacua. Moreover, they can be directly implemented in the framework of computer-aided search and can be easily combined with other phenomenological constraints.

We explicitly derive these constraints and
apply our results in the investigation of  a big class of Pati--Salam vacua, $G=SU(4)\times{SU(2)}_L\times{SU(2)}_R$, consisting of $10^{16}$ models.  
The contribution of Yukawa coupling selection criteria is critical in this 
case as it allows, when combined with other phenomenological requirements,  to go beyond statistical sampling utilised so far in the examination of these vacua and to fully derive  all consistent models. An exhaustive scan yields approximately $10^7$ models that satisfy an extensive set of phenomenological requirements. These vacua will be further 
analysed in a future publication.

Our results can also be applied to the study of other classes of $Z_2\times Z_2$ vacua in the framework of the Free Fermionic Formulation, as the recently analysed flipped models, $G=SU(5)\times U(1)$, \cite{flipped} or models based on other $SO(10)$ subgroups.
The analysis can also be extended to include other fermion couplings starting from the 
bottom quark coupling. In the case of Pati--Salam models, considered here, the extension is trivial since both the top and bottom Yukawa couplings arise from the same superpotenial term.  For flipped-$SU(5)$ models the bottom Yukawa coupling may involve one untwisted Higgs
doublet. The relevant correlator has been calculated in section 2. It is thus  straightforward to include the bottom coupling in the analysis.
Yukawa couplings of the lighter generations, that potentially arise from
 higher order nonrenormalisable terms in the effective superpotential can also be 
 analysed in a similar way but the computation is far more intricate as it requires the investigation of correlators involving also the singlet states that might develop vacuum expectation values. Yukawa coupling related criteria can  thus be a valuable guide in the quest for phenomenologically viable models in the immense space of string vacua.

\section*{Acknowledgements}
This research has been co-financed by the European Union
(European Social Fund - ESF) and Greek national funds through the Operational Program ``Education and Lifelong Learning" of the National Strategic Reference Frame-work (NSRF) - Research Funding Program: ``ARISTEIA". Investing in the society of knowledge through the European Social Fund. The author would like to thank CERN Theory Division for the 
hospitality during the  preparation of this work.

\end{document}